# A scientific inquiry into modern art*

Mikhail Simkin

“He suggested I play golf, but finally agreed to give me something that, he said, "would really work"; and going to a cabinet, he produced a vial of violet-blue capsules banded with dark purple at one end, which, he said, had just been placed on the market and were intended not for neurotics whom a draft of water could calm if properly administered, but only for great sleepless artists who had to die for a few hours in order to live for centuries.”

*-- Vladimir Nabokov, Lolita*

Are the sleepless artists really great, or merely properly administered?

To check this I wrote the “True art or a fake?” quiz [1]. It consists of a dozen pictures. Some of them are True Masterpieces of Abstract Art, created by Immortal Artists. The rest are ridiculous fakes, produced by myself. The takers are to tell which is which.

Apart from the scores, which are automatically recorded, occasionally I get feedback. This note had arrived from a Cornell University professor: “I recognized that No. 2 was like a Mondrian, but it seemed to lack the sense of balance which good modern art is supposed to have”. Apparently, Mondrian’s art loses balance when his heavy-weight name is detached from it. Even art critics are not sure that they can tell true art from fake: “I got 92%, which is a relief since I write about art.” It is thus not surprising that sometimes the quiz provokes angry reaction. One New York artist responded with the following utterance: “Go [profanity] yourself and your [profanity] academic quizzzzzzzzzzzzzzzzzz.” As if in response to this attempt of intimidation, one of my readers wrote: "Dear Mr. Simkin, just continue with this". And I did.

The distribution of the scores received by over fifty thousands quiz-takers† is shown in Figure 1.The average score is 7.91 out of 12 or 65.9% correct.

Our respondents did poorly on the test. But could this be because they are a bunch of philistines, interested only in the material side of life and vulgar in taste? I don’t know them personally, as the testing was done over the internet. However, the quizzing software [2] records taker’s IP address. From it one can infer where their computer was. This enabled me to select the test scores received by people who downloaded the quiz from elite locations. For the analysis I chose Ivy League schools and Oxbridge (if not for any other reason, than because I did time in those places). The average elite score is 8.5 out of 12 or 71% correct. Figure 2 shows the distribution of the scores received by 143 chosen quiz-takers, and Table 1 shows their distribution by elite schools. When comparing the averages and the distributions of scores, shown in Figures 1 and 2, we see that there is not much difference between the elite and the crowd.

---

† There were 71,456 test results in the database. However a number of people took several shots at the quiz. I cleaned the data from the second-attempt scores by selecting only the first score from each IP address. This reduced the number of test scores to 61,121. Next I cleaned the data from the results, where one or more questions were skipped. This reduced the number of the test scores to 56,020.

Although the performance of our quiz-takers is poor, it is better than what one can do by random guessing. The latter would produce a symmetric distribution of scores centered at six correct answers, and the average score of 50%. Does this mean that there is a small, but perceptible difference in quality between abstract masterpieces and fakes? Probably not. From the feedback, I know that many quiz takers had previously seen some or all of the masterpieces, *identified as such*. For example, one of the respondents wrote: "I gave this test to my oldest son who is teaching sculpture at The Finnish Art Academy. Much to my chagrin, he could not only separate the art from the chaff, but also name all the artists." Another successful strategy, used by many high-scores, is well summarized in this note: "I got 100% in your quiz. Why? Because I could tell immediately which were created on a computer and which were created on canvas".

Apart from explicit mentions in feedback, the fact that many quiz takers had previously seen some of the masterpieces can be directly inferred from the distribution of test scores. The average score of 65.9% means that an average image was identified correctly as true art or fake in 65.9% of the cases. An interesting thing is that this splits unevenly between masterpieces and fakes. An average masterpiece was correctly identified as such in $p_m = 67.5\%$ of the cases, while an average fake was correctly identified as such in only $p_f = 64.3\%$. The standard errors of both $p_m$ and $p_f$ are 0.1%, so the difference $p_m - p_f = = 3.2\%$ is statistically significant. The obvious explanation of this observation is that some of the quiz-takers had previously seen some of the masterpieces. If someone had seen the image in an art gallery or in an album he will tick it as true art. If he didn't see it before, he will have to use other criteria, for example, whether image was created on a computer or on canvas. As our quiz-takers haven't previously seen any of the fakes, the percentage of correct identification of a fake should be equal to the percentage of the correct identification of a masterpiece in the case that the quiz-taker hasn't seen it before. This can be used to estimate the fraction, $f$, of the masterpieces, used in the quiz, that were previously seen by the quiz-takers. The probability, $p_m$ , to identify a masterpiece correctly can be splits in two terms. If the taker has seen the masterpiece before (what happens with probability $f$), he identifies it correctly with probability 1. If he hasn't seen the masterpiece before (what happens with probability $1-f$), he identifies it correctly with probability $p_f$. Thus: $p_m = f + (1 - f)\, p_f$. From this follows: $f = (p_m - p_f) / (1 - p_f) = 9\%$.

Although the quiz results are biased in favor of masterpieces, I'll take them at face value to quantify the difference in quality between the images. Table 2 shows for each picture the fraction of quiz-takers who selected it as "true art". The top-rated painting was ticked "true" by 80% of people, and the bottom-rated by only 15%. What does this say about the difference in intrinsic quality?

A hundred years ago psychologist F.M. Urban conducted the classic study of *just perceptible differences* [3]. He asked the subjects of his experiment to compare a hundred-gram weight with a set of different weights. When two weights were very close the subject's judgment was poor. However, statistically, the lighter weight was perceived to be heavier in less than fifty percent of the cases. For example, 92, 100, and 104 gram were perceived heavier than 100 gram, in 10%, 50%, and 84% of trials correspondingly. I defined the "weight" of a hypothetical painting, which is selected as true art by 50% of quiz takers as 100 gram. Afterwards, by interpolation of Urban's data, I inferred the "weights" of the paintings, used in the quiz (see Figure 3). They are given in the rightmost column of Table 2.

The difference in weight between the lightest (93.8 g) and the heaviest (103.4 g) pictures is ten percent. For comparison, in the sport of weight lifting men weighting between 94 and 105 kg belong to the same weight category [4]. I conclude that all pictures in the quiz, when judged by their intrinsic qualities, fall into the *same weight category*. The only difference between masterpieces and fakes is in *heavy-weight names* attached to the masterpieces.

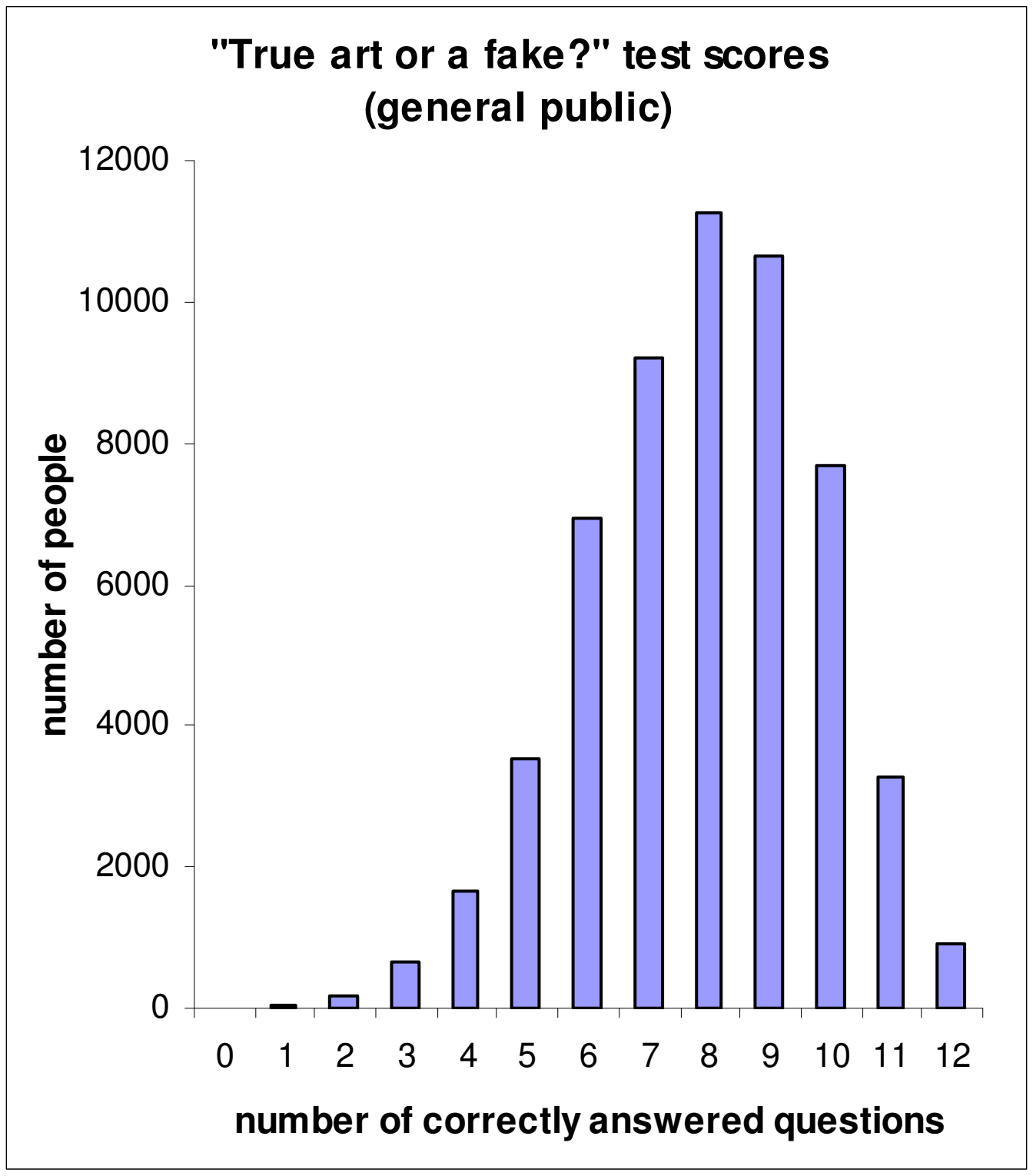

**Figure 1.** The histogram of the test scores, earned by 56,020 quiz-takers. The average score is 7.91out of 12 or 65.9% correct.

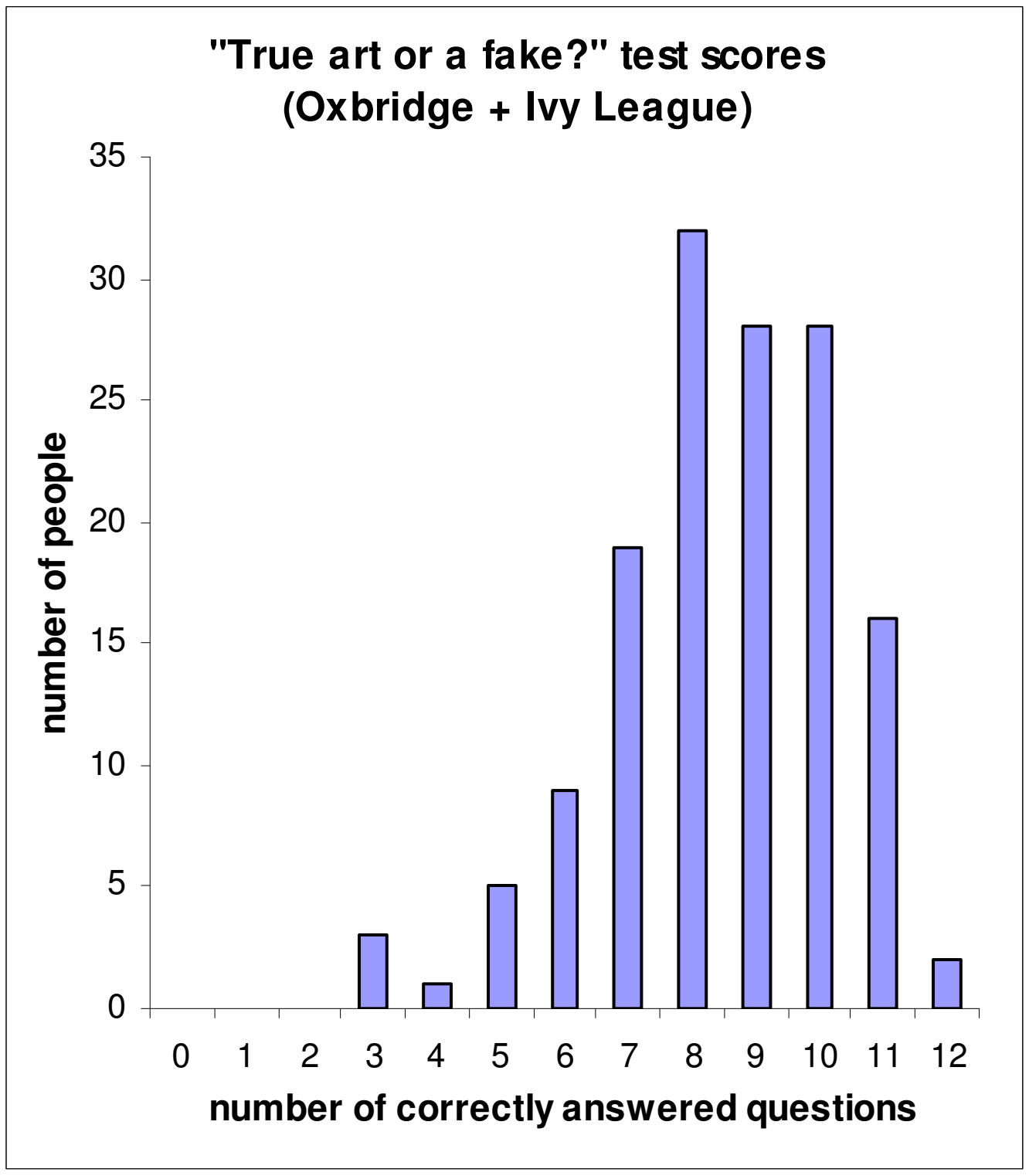

**Figure 2.** The histogram of the test scores, earned by 143 chosen quiz-takers. The distribution of the chosen by elite schools is given in Table 1. The average elite score is 8.5 out of 12 or 71% correct.

**Table 1** The distribution of the chosen quiz-takers (whose scores are shown in Figure 2) by elite universities.

| Elite University | Number of quiz-takers | Score | | |
|---|---|---|---|---|
| | | minimum | maximum | average |
| Brown | 5 | 7 | 11 | 8.8 |
| Cambridge | 24 | 3 | 12 | 8.4 |
| Columbia | 28 | 5 | 11 | 8.3 |
| Cornell | 7 | 5 | 11 | 8.0 |
| Dartmouth | 6 | 7 | 9 | 8.0 |
| Harvard | 22 | 5 | 11 | 8.8 |
| Oxford | 24 | 3 | 10 | 8.2 |
| Princeton | 7 | 7 | 11 | 8.6 |
| Penn | 8 | 7 | 10 | 8.5 |
| Yale | 12 | 8 | 12 | 9.3 |
| Total | 143 | 3 | 12 | 8.5 |

**Table 2**. For each picture, the fraction of quiz takers, who selected it as true art, is shown alongside with picture's "weight", determined by comparison with Urban's data. You can have a look at the pictures themselves on the quiz's webpage [1].

| picture number | artist | percent of selection as true art | artistic weight (in artistic grams) |
|---|---|---|---|
| 9 | Kandinsky | 0.79 | 103.4 |
| 2 | Mondrian | 0.76 | 102.9 |
| 8 | Rothko | 0.75 | 102.7 |
| 12 | Albers | 0.67 | 101.9 |
| 4 | Malevich | 0.61 | 101.2 |
| 6 | fake | 0.58 | 100.8 |
| 1 | Klee | 0.46 | 99.4 |
| 10 | fake | 0.39 | 98.0 |
| 11 | fake | 0.37 | 97.7 |
| 3 | fake | 0.36 | 97.5 |
| 7 | fake | 0.28 | 95.9 |
| 5 | fake | 0.17 | 93.8 |

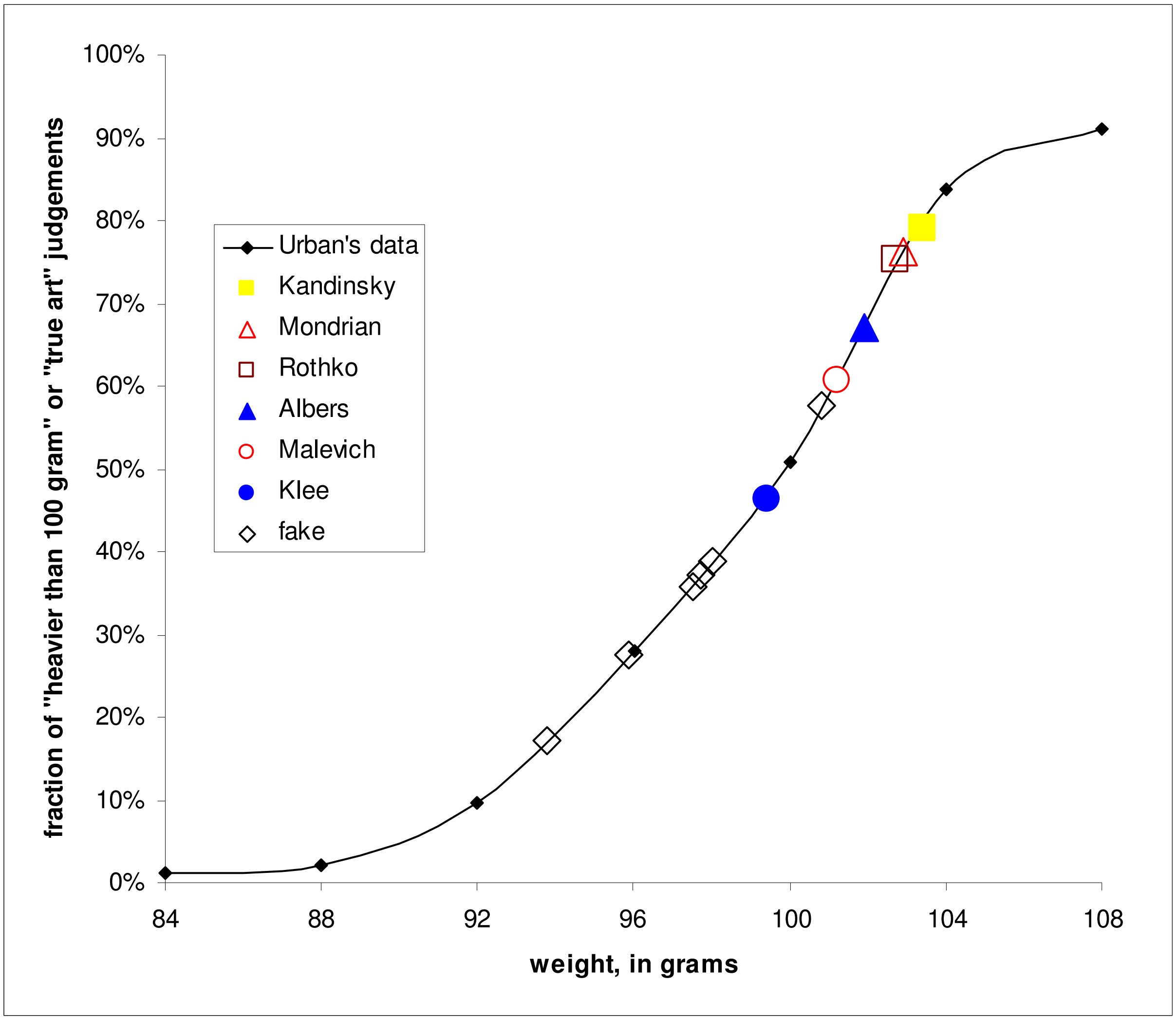

**Figure 3.** Small rhombs represent Urban's data on the fraction of "heavier" judgments for different weights (with the control weight of 100 gram). The line is the interpolation of that data. The larger symbols represent the pictures used in the quiz. Picture's "weight" was adjusted so that the symbols fall on the interpolation line. You can have a look at the pictures themselves on the quiz's webpage [1].